\documentclass{article}
\usepackage{amsmath}
\usepackage{amssymb}
\usepackage{cite}

\begin{document}

\newcommand{\bear}{\begin{eqnarray}}
\newcommand{\eear}{\end{eqnarray}}
\newcommand{\be}{\begin{equation}}
\newcommand{\ee}{\end{equation}}
\newcommand{\beqn}{\begin{eqnarray}}
\newcommand{\eeqn}{\end{eqnarray}}
\newcommand{\beqnn}{\begin{eqnarray*}}
\newcommand{\eeqnn}{\end{eqnarray*}}

\def\vep{\varepsilon}
\def\vf{\varphi}

\thispagestyle{plain}

\label{sh}

\begin{center} {\Large \bf
EXCITATION OF A MOVING OSCILLATOR}

\end{center}

\begin{center} {\bf
Viktor~V.~Dodonov  
}\end{center}

\begin{center}

{\it 
Institute of Physics and International Center for Physics,
University of Brasilia \\
70910-900 Brasilia, Federal District, Brazil}

$^*$E-mail:~~~vdodonov\,@\,fis.unb.br\\
\end{center}

\begin{abstract}\noindent

We calculate transition amplitudes and probabilities between the coherent and Fock states of a quantum harmonic
oscillator with a moving center for an arbitrary law of motion. These quantities
are determined by the Fourier transform of the moving center acceleration. 
In the case of a constant acceleration, the probabilities oscillate with the oscillator frequency, 
so that no excitation occurs after every period. 
Examples of oscillating and rotating motion of the harmonic trap center are
considered too. Estimations show that the effect of excitation of vibration states
due to the motion of the harmonic trap center can be observed in available atomic traps.

\end{abstract}



\section{Introduction} 

The problem of evolution of non-relativistic quantum packets in moving potentials (moving traps) with fixed shapes
attracted attention of many authors for the past five decades.
In particular, moving rectangular potential wells or barriers in one dimension were considered in 
\cite{Pimpale91,Lee02,Chiofalo03,Shegelski13,Dimeo14}.
The moving  Eckart potentials were studied  in \cite{Ge96,Miyashita07}.
A moving one-dimensional rigid box with impenetrable walls was studied in \cite{Hron81,Beauchard06}.
The moving delta-potential in one dimension was discussed in \cite{Lee02,Pimpale91f,Cheng93,Pimpale04,Granot09}.
Two moving delta-potentials in one dimension were investigated in 
\cite{Breit65,Herling65,Nishida65,Zhdanov74,Solov76,Dappen77,Danared84,Scheitler90,Manko98,Chikhachev05tmf,Chikhachev05}.
Generalizations to three dimensions were made in \cite{Solov76,Chikhachev05tmf,Chikhachev05}, while 
a constant harmonic oscillator potential was added in \cite{Chikhachev05}.
The Dirac and Klein--Gordon equations for the moving step potential were solved in \cite{Hamil16}.

A non-relativistic quantum harmonic oscillator with a moving center described by means of the
time-dependent Schr\"odinger equation with the Hamiltonian
\be
H= p^2/(2M) + M\omega^2[x-b(t)]^2/2 ,
\label{Ham}
\ee
was studied with different purposes in \cite{Duru89,Hurley11}. Several authors 
considered Hamiltonian (\ref{Ham}) in connection with the problem of optimal non-adiabatic transport of traps for ultracold atoms
without exciting  their vibrational quantum states (heating) \cite{Reichle06,Couvert08,Torron11,Guery14}.
The subject of the present article is the general problem of {\em excitation\/} of a {\em moving\/} oscillator,
which seems to be not considered until now in full. The search for non-exciting trajectories of the moving
center is the special case of this general problem.

\section{Transition probabilities to the instantaneous Fock states of the moving oscillator}

Hamiltonian (\ref{Ham}) can be written as the Hamiltonian of a  forced
harmonic oscillator 
\be
H= p^2/(2M) + M\omega^2x^2/2 -f(t) x + f^2(t)/(2 M\omega^2),
\label{Hamf}
\ee
with the ``effective force'' $f(t) = M\omega^2 b(t)$.
The dynamics of a quantum oscillator with Hamiltonian (\ref{Hamf}) was the subject of numerous studies.
In particular, this problem attracted attention of many researchers in 1950s-1960s
\cite{Bart49,Feynman50,Ludwig51,Feynman51,Glauber51,Husimi53,Schwinger53,Kerner58,Fuller63,Scarfone64,%
Carr65,Gilbey66,Ninan69,Merz70,Zeld70,MM70}. 
For detailed reviews see, e.g., \cite{DM183-2,Popov07}.
One of the most important results was a nice formula for the probabilities of transitions between the initial Fock state
$|m\rangle$ at $t=0$ and the Fock state $|n\rangle$ at the instant $t$
\be
P_{mn} =\frac{\mu!}{\nu!} \gamma^{|m-n|} e^{-\gamma}\left[L_{\mu}^{(|m-n|)}(\gamma)\right]^2, \quad
\begin{array}{l}
\mu =\mbox{min}(m,n), \\ \nu =\mbox{max}(m,n),
\end{array}
\label{Pmn}
\ee
where $L_n^{(\alpha)}(x)$ is the associated Laguerre polynomial (defined according to \cite{Bateman}).
Parameter $\gamma$ is related to the Fourier transform of the external force:
\be
\gamma =(2M\hbar \omega)^{-1}\left|\int_0^t f(\tau) e^{i\omega \tau}d\tau\right|^2.
\label{gam}
\ee
Formula (\ref{Pmn}) was found for the first time by Ludwig \cite{Ludwig51}, but his work was practically unknown,
since it was written in German. Two years later the same result was given by Schwinger \cite{Schwinger53} 
(for $t=\infty$ and, strictly speaking, for the field oscillators excited by a classical current). 
The special case of (\ref{Pmn})  for $m=0$, 
\be
P_{0n}= e^{-\gamma} \gamma^n /n!,
\label{P0n}
\ee
was derived in \cite{Bart49,Feynman51,Glauber51}, but the presence of the Laguerre polynomial was not detected 
in that papers.
Husimi \cite{Husimi53} used Charlier polynomials instead of Laguerre ones.
Later, different derivations (sometimes rather long and complicated) of (\ref{Pmn}) were given in
\cite{Kerner58,Fuller63,Scarfone64,Carr65,Gilbey66,Ninan69,Merz70}.

However, applying formula (\ref{Pmn}) with parameter $\gamma$ defined by (\ref{gam}) 
with $f(t) = M\omega^2 b(t)$ to the case of moving oscillator, one arrives at strange results.
For example,  in the special case of uniformly moving center with $b(t)=vt$ 
(assuming that the center position at $t=0$ was $x=0$) one obtains
\be
\gamma(t) = \frac{Mv^2}{2\hbar\omega}\left[(\omega t)^2 +4\sin^2(\omega t/2) - 2\omega t \sin(\omega t)\right].
\label{gamv}
\ee 
This formula predicts unlimited excitations of the oscillator energy levels. 
The reason of the controversy becomes clear, if one looks at solutions of the classical equations of motion
for Hamiltonian (\ref{Ham}) (they coincide with quantum solutions for the average values). In the simplest
case of zero initial conditions, these solutions read
\be
x(t) = vt -v\sin(\omega t)/\omega, \quad d{x}/t = v - v\cos(\omega t).
\label{xclas}
\ee
Therefore, the first term in (\ref{gamv}) (proportional to $t^2$) is nothing but the potential energy of the moving
oscillator center (normalized by the energy quantum $\hbar\omega$). It appears because
 formula (\ref{Pmn}) describes transitions
between the Fock states defined for the oscillator with a {\em fixed\/} center at $x=0$, while we are interested
in the situation with the {\em moving\/} center. The amplitude of real oscillations in the case concerned is proportional
to the constant velocity $v$ of the moving center, so that the excitation probabilities (related to the energy
of oscillations) must be also time-independent ones.

One of the simplest ways to obtain the correct result is to follow the scheme used in \cite{MM70,DM183-2}, 
based on the concept of coherent states. 
The time dependent coherent state, satisfying the Schr\"odinger equation
with Hamiltonian (\ref{Hamf}) and coinciding with the usual (unforced) coherent state at $t=0$,
has the following form in the coordinate representation (see, e.g., \cite{DM183-2}):
\beqn
\langle x|\alpha\rangle_t &=& \left(\frac{M\omega}{\pi\hbar}\right)^{1/4} \exp\left\{
-\frac{M\omega}{2\hbar} x^2 + \sqrt{\frac{2M\omega}{\hbar}}\, x e^{-i\omega t}[\alpha -\delta(t)] 
+ \alpha \left[\delta(t) e^{-2i\omega t} +\delta^*(t)\right]
\right. \nonumber \\
&& \left.  
-\frac{i\omega t}{2} -\,\frac{\alpha^2}{2}e^{-2i\omega t}  -\,\frac{|\alpha|^2}{2}
 +i\omega \int_0^t d\,\tau \delta^2(\tau)e^{-2i\omega \tau} 
 - \frac{i}{2 M\omega^2\hbar} \int_0^t f^2(\tau)\,d\,\tau 
 \right\},
\label{coh}
\eeqn
where
\be
\delta(t) = -i(2M\hbar\omega)^{-1/2} \int_0^t f(\tau) e^{i\omega \tau}d\tau,
\qquad |\delta(t)|^2 = \gamma.
\label{del}
\ee
One has to compare this function with the coherent state $|\beta\rangle$ of the unforced oscillator (with $\delta=0$),
considered  with respect to the {\em moving\/} center $b(t)$
(not with respect to the initial oscillator center at $x=0$). Applying the {\em extended\/} Galilean transform
\cite{Rosen72,Greenberger79}, the corresponding wave function can be written as
\beqn
\langle x|\beta\rangle_t &=& \left(\frac{M\omega}{\pi\hbar}\right)^{1/4} \exp\left\{
-\frac{M\omega}{2\hbar} [x-b(t)]^2 +
\sqrt{\frac{2M\omega}{\hbar}}\, [x -b(t)] e^{-i\omega t}\beta +\frac{iM\dot{b}(t)}{\hbar} x
\right. \nonumber \\
&& \left.  -\frac{i\omega t}{2}
-\,\frac{\beta^2}{2}e^{-2i\omega t}  -\,\frac{|\beta|^2}{2}
 - \frac{iM}{2 \hbar} \int_0^t \dot{b}^2(\tau)\,d\,\tau 
 \right\},
\label{cohG}
\eeqn
where $\dot{b} \equiv db/dt$. The transition amplitude between coherent states $\langle \beta|\alpha\rangle$
is given by the Gaussian integral. After some algebra (in particular, using several integrations by parts
in the integrals containing functions $\delta^2(\tau)$ and $\dot{b}^2(\tau)$),
one can arrive at the following simple expression:
\be
\langle \beta|\alpha\rangle = \exp\left[\alpha\beta^* +\alpha u(t) -\beta^* u^*(t)
-\left(|\alpha|^2 +|\beta|^2 \right)/2
-|u(t)|^2/2 -i\phi(t) \right],
\label{Tab}
\ee
where
\be
u(t) = -i \sqrt{\frac{M}{2\hbar\omega}} \int_0^t \ddot{b}(\tau) e^{-i\omega\tau}\,d\,\tau,
\label{ut}
\ee
\be
\phi(t) = \int_0^t d\tau \left\{\mbox{Im}\left[\dot{u}(\tau)u^*(\tau)\right]
+M b(\tau)\ddot{b}(\tau)/\hbar \right\}.
\label{phi}
\ee
The simple expression for the phase (\ref{phi}) was obtained under the conditions $b(0)=\dot{b}(0)=0$
(used during some integrations by parts), but these conditions were not used for the derivation of
function $u(t)$ (\ref{ut}).

Using the standard expansions of coherent states over the (time-dependent) Fock states,
\[
|\alpha\rangle_t = e^{-|\alpha|^2/2} \sum_{m=0}^{\infty} \frac{\alpha^m}{\sqrt{m!}}|m\rangle_t,
\quad
|\beta\rangle_t = e^{-|\beta|^2/2} \sum_{n=0}^{\infty} \frac{\beta^n}{\sqrt{n!}}|n\rangle_t,
\]
one can see that function (\ref{Tab}) is the generating function for the transition amplitudes $A_{mn}$ between
the evolved initial states $|m\rangle_t$ and moving unforced states $|n\rangle_t$:
\[
\langle \beta|\alpha\rangle  = \exp\left[-\left(|\alpha|^2 +|\beta|^2\right)/2\right]
\sum_{m,n=0}^{\infty} \frac{\alpha^m \beta^{*n}}{\sqrt{m! n! }}A_{mn}.
\]
Taking into account the generating function of the associated Laguerre polynomials 
(proofs can be found, e.g., in \cite{DM183-2,CahGla1})
\be
\exp\left(\xi x +\eta y + \xi\eta\right) = \sum_{m,n=0}^{\infty} \frac{\xi^n \eta^m}{n!} 
x^{n-m} L_m^{(n-m)}\left(-xy\right),
\label{Lagg}
\ee
one arrives finally to formula (\ref{Pmn}) for the transition probabilities $P_{mn} = \left|A_{mn}\right|^2$,
but with $\gamma = |u(t)|^2$. This is the first main result of this article.
Formally, it could be obtained by putting  the force of inertia $f_i = -M \ddot{b}$ in formula (\ref{gam}).
But our goal was to provide the strict proof, using the extended Galilean transform.
Our second goal was to analyze whether the effects of excitation due to the motion of the center
of a harmonic potential (harmonic trap) could be observed under realistic experimental conditions.
The examples of the next section show that the answer is positive.

\section{Examples and numerical estimations}

If $\ddot{b}=a = const$, then
\be
u(t) = \frac{a \sqrt{M}}{2\hbar\omega^3}\left(e^{-i\omega t} -1 \right), \;\;\;
|u(t)|^2 = \frac{2M a^2}{\hbar\omega^3}\sin^2\left(\frac{\omega t}{2}\right),
\label{ua}
\ee
so that a constant acceleration results in {\em periodic\/} excitations, and the oscillator returns to its
initial internal state (in the uniformly accelerated frame) after each period $2\pi/\omega$.

If the oscillator center is accelerated during a short time interval $T_a \ll 2\pi/\omega$, 
then $|u(t)|^2 = Mv^2/(2\hbar\omega) \equiv G$ for $t>T_a$, 
independently of the concrete form of function $\ddot{b}(t)$.
Here $v$ is the final constant velocity of the  moving center. 
For example, if the center of a harmonic trap for ultracold atoms with frequency $\omega \sim 100$\,s$^{-1}$ 
(this is a realistic value \cite{Leggett01}) almost instantly starts to move with the constant velocity
$1\,$mm/s, then $G \sim 5$ for atoms with mass $M\sim 10^{-22}\,$g, which means a significant degree
of excitation. If the uniformly moving center suddenly stops later at instant $T$, then 
$|u(T)|^2 = 4G\sin^2(\omega T/2)$.

For the periodic motion of center, $b(t)= R[1-\cos(\Omega t)]$ (satisfying the conditions $b(0)=\dot{b}(0)=0$),
  one obtains
\be
 |u(t)|^2 =G\,\Omega^2\left[\frac{\sin^2(\omega_- t/2)}{\omega_-^2}
+ \frac{\sin^2(\omega_+ t/2)}{\omega_+^2}
+ \frac{2\cos(\Omega t)\sin(\omega_- t/2)\sin(\omega_+ t/2)}{\omega_- \omega_+}
\right]. 
\label{uosc}
\ee
Here, $\omega_{\pm} = \omega \pm \Omega$ and $G = M(R \Omega)^2/(2\hbar\omega)$.
If the center returns to its initial position after $s$ oscillations at $t=2\pi s/\Omega$, then
\be
 |u_s|^2 = \frac{4G(\Omega\omega)^2}{\left(\Omega^2 -\omega^2\right)^2}\sin^2(s\pi\omega/\Omega).
\label{uosc-s}
\ee
The same value can be obtained for the parameter $|\delta_s|^2$ calculated according to Eq. (\ref{del}),
since the centers of moving and fixed oscillators coincide at this specific instant.
One can see that $|u_s|^2 \ll G$ if $\omega \ll \Omega$ or $\omega \gg \Omega$. 
But this parameter can be quite big in the resonance case $\omega=\Omega$ (as one can expect):
$\left|u_s^{(res)}\right|^2 = G(\pi s)^2$.

For the two-dimensional oscillator moving in the $xy$ plane, the coherent states and transition amplitudes between them, being
the Gaussian exponential functions, are obvious products of functions (\ref{coh}), (\ref{cohG}) and (\ref{Tab}).
Consequently, transition probabilities between the selected initial $|m_x, m_y\rangle$ and final $|n_x, n_y\rangle$ Fock states
are given by products of functions (\ref{Pmn}), dependent on parameters  $|u_x(t)|^2$ and $|u_y(t)|^2$ determined by the
moving center coordinates $b_x(t)$ and $b_y(t)$.
If the oscillator is isotropic, then the total transition probabilities between the energy eigenstates
(taking into account their degeneracy)
must depend on the only parameter $|{\bf u}|^2 =\left|u_x\right|^2 + \left|u_y\right|^2$, due to  the
rotational symmetry. 
This statement can be easily confirmed for the transitions from
the vacuum state $|0,0\rangle$ to all states $|n_x,n_y\rangle$ with $n_x+n_y = n =const$,
\be
{\cal P}_{0n} = \exp\left[-\left(\left|u_x\right|^2 + \left|u_y\right|^2\right)\right] \sum_{n_x+n_y = n}
\frac{\left|u_x\right|^{2 n_x}\left|u_y\right|^{2 n_y}}{n_x ! n_y !} 
=
\exp\left(-|{\bf u}|^2\right) \frac{|{\bf u}|^{2n}}{n!}.
\label{Pon2}
\ee
Here, the known binomial formula was used. The same result (\ref{Pon2}) is true for the $N$-dimensional
isotropic harmonic oscillator (as a consequence of the multinomial formula), but only for transitions
from the initial ground state.

One could suppose that a general formula
for the total transition probability ${\cal P}_{mn}$ between the $m$-th and $n$-th degenerate energy levels
is given by the same formula (\ref{Pmn}) with $\gamma$ replaced by $|{\bf u}|^2 \equiv w$. 
Unfortunately, it is not the case.
One of simple examples is the transition probability between all the first and second excited
energy levels of the two-dimensional isotropic oscillator (the notation below seems to be clear):
\[
{\cal P}_{12} = P_{01,02} + P_{01,11} + P_{01,20} + P_{10,02} + P_{10,11} + P_{10,20}.
\]
Using explicit expressions for the associated Laguerre polynomials $L_1^{(\alpha)}(x) = 1 +\alpha - x$  \cite{Bateman},
one obtains (writing here $\left|u_x\right|^2 =\xi$ and $\left|u_y\right|^2 =\eta$, so that 
$\xi+\eta=  w$)
\beqnn
{\cal P}_{12} &=& e^{-\xi-\eta}\left[\frac{\eta}{2}(2-\eta)^2 +\xi(1-\eta)^2 +\frac{\xi^2}{2}\eta 
 +
\xi \frac{\eta^2}{2} +(1-\xi)^2\eta +\frac{\xi}{2}(2-\xi)^2\right]
\nonumber \\
&=& \frac{w}{2}e^{-w}\left(6 - 4w +w^2\right)
= \frac{w}{2}e^{-w} \left\{2 +\left[L_1^{(1)}(w)\right]^2 \right\}.
\eeqnn
 Similarly, ${\cal P}_{11} =  e^{-w} \left\{ 1 +\left[L_1^{(0)}(w)\right]^2 \right\}$.

Now we consider the motion of the center of the isotropic oscillator potential along a circle:
$b_x(t) = R\{1-\cos[\vf(t)]\}$, $b_y(t) = R\sin[\vf(t)]$ 
(this is an example of rotating traps \cite{Fetter09,Palm16,Lizu19}).
Function $\vf(t)$ must satisfy the condition 
$\dot\vf(0)=0$, due to the requirements $b_x(0)=b_y(0)=\dot{b}_x(0)=\dot{b}_y(0)=0$. 
If $\dot\vf(t)$ grows rapidly (during the time interval $T_a$ much smaller than $2\pi/\omega$) to the constant value $\Omega$,
 one can put $\vf(t)=\Omega t$ in the integral for $u_x(t)$. Consequently, function $|u_x(t)|^2$ is given again by
Eq. (\ref{uosc}).
But calculating $u_y(t)$, one must take into account rapid jumps of the velocity $\dot{b}_y$ from zero to $R\Omega$ at the
initial short transient interval and its fast return to zero value before  $t_s=2\pi s/\Omega$,
if the moving center rapidly stops at the initial point after $s$ complete revolutions.
Finally, one can obtain the following expression for
the excitation parameter $w_s = |u_x(t_s)|^2 + |u_y(t_s)|^2$:
\be
w_s  = \frac{2M(R\Omega\omega)^2\left(\omega^2 +\Omega^2\right)}{\hbar\omega\left(\omega^2 -\Omega^2\right)^2}
\sin^2(s\pi\omega/\Omega).
\label{ws}
\ee
For the slowly rotating ($\Omega \ll \omega$) harmonic trap center one has
\be
w_s =G\sin^2(s\pi\omega/\Omega), \quad G= 2MR^2\Omega^2/(\hbar\omega).
\ee
For example, if the center of a harmonic trap  with $\omega \sim 100$\,s$^{-1}$ 
 slowly rotates with frequency $\Omega \sim 10^{-2}$\,s$^{-1}$ (one rotation for ten minutes) along a circle
with radius $R=10$\,cm, then $G \sim 20$ for $M \sim 10^{-22}$\,g.
Consequently, strong excitations of the vibration modes inside the trap can be achieved, although the concrete
value of the excitation parameter $w_s$ is very sensitive to the value of the big ratio $\omega/\Omega$.
Perhaps, this result could be interesting from the point of view of experiments with ultracold gases in
harmonic traps.

\section*{Acknowledgment}

The author acknowledges a partial support of the Brazilian agency CNPq.

\end{document}